\documentstyle[12pt, a4]{article}
\begin{document}
\author{O. \"{O}zer and B. G\"{o}n\"{u}l
\and Department of Engineering Physics,Faculty of Engineering,
\and University of Gaziantep, 27310 Gaziantep-T\"{u}rkiye}
\title{New exact treatment of the perturbed Coulomb interactions}
\date{}
\maketitle
\begin{abstract}
A novel method for the exact solvability of quantum systems is
discussed and used to obtain closed analytical expressions in
arbitrary dimensions for the exact solutions of the hydrogenic
atom in the external potential $\Delta V(r)=br+cr^{2}$, which
is based on the recently introduced supersymmetric perturbation
theory.
\end{abstract}

One of the challenging problems in nonrelativistic quantum mechanics
is to find exact solutions to the Schr\"{o}dinger equation for
potentials that are used in different fields of physics. In particular,
the perturbed Coulomb potentials represent simplified models
of many situations found in atomic, molecular, condensed matter
and particle physics. There has been much interest in obtaining
analytical solutions of such potentials in arbitrary dimensions.
These problems have been studied for years and a general solution
has not yet been found.

Such class of potentials,
\begin{equation}
V(r)=-a/r +br+cr^{2}~,
\end{equation}
are possible candidates for the quarkonium potential as has been
indicated by the quarkonium spectroscopy. In the special case
of $c=0$ and $b>0$ such potentials reduce to the well known charmonium potential.
Apart from its relevance in heavy quarkonium spectroscopy, this
class of potentials with $c=0$ has important applications in atomic
physics. The Stark effect in a hydrogen atom in one dimension is
given exactly by the charmonium like potential ($b$ being the
electric field parameter). The more general class of these potentials
with $c>0$ is also relevant in atomic physics. This could be interpreted
as the potential seen by an electron of an atom exposed to a
suitable admixture of electric and magnetic fields. In addition,
nuclei in the presence of an electron background form a system
which is important for condensed matter physics and for laboratory
and stellar plasmas. The potential between two nuclei embedded
in such a plasma is approximately Coulomb plus harmonic oscillator,
which corresponds to $b=0$ in (1).

As the exact form of such interactions are being unknown to a
great extent, it is thus desirable to study the general analytical
properties of a large class of potentials in (1). In connection
with this, the analyticity of the energy levels for these kind
of potentials was investigated rigorously by many authors using
different theories \cite{killingbeck131978}-\cite{alberg72001} in
relation to their potential applications in spectroscopic problems.

In this letter, we introduce an alternative, simple formalism
for an algebraic solution of the Schr\"{o}dinger equation with
the perturbed Coulomb potential and find exact solutions
in $N$-dimesional space. The new formalism is based on the supersymmetric
quantum mechanics and uses the spirit of perturbation theory,
in which there is no room for the drawbacks encountered in the
calculations with the perturbation theories available in the
literature. The work introduced in this letter also clarifies,
within the powerful frame of the present formalism, that the
potential in (1) is in fact an exactly solvable shape invariant
potential unlike the claim in a recent work \cite{chaudhuri18501995} where the authors
stated that the supersymmetric quantum mechanics yields exact
solutions for a single state only for the perturbed Coulomb potential
in (1) which was treated as a quasi-exactly solvable potential.

Let us now introduce the formalism. The reduced radial wave Schr\"{o}dinger
equation for a spherically symmetric potential in $N$-dimensional space reads
\begin{equation}
\frac{\hbar^{2}}{2m}\frac{\Psi''_{n}(r)}{\Psi_{n}(r)}=V(r)-E_{n}~,~
V(r)=V_{ES}(r)+\Delta V(r)~,~
V_{ES}(r)=V_{0}(r)+\frac{\Lambda\left(\Lambda+1\right)\hbar^{2}}{2mr^{2}}~,
\end{equation}
where $V_{0} $ is one of the exactly solvable potential, such as the Coulomb
and harmonic oscillator potentials used through the present work, $n=0,1,2,\ldots$
being the radial quantum number, and $\Lambda =( M-3)/2$ with $M=N+2\ell $. We
see that the radial Schr\"{o}dinger equation in $N$ dimensions
has the same form as the three-dimensional one. Consequently,
given that the potential has the same form in any dimension,
the solution in three dimensions can be used to obtain the solution
in arbitrary dimensions simply by the use of the substitution
$\ell \rightarrow \Lambda $. In the above equation, $\Delta V$ stands
for the small perturbation $\left( =br+cr^{2}\right)$ assuming
the potential parameter $a$ is large when compared to other coupling
parameters $b$ and $c$ in (1) with the consideration of
perturbed Coulomb interactions while for the case of large $c$ and
relatively small $a$ and $b$ the perturbing potential involves
$\Delta V=-a/r +br$ leading to a perturbed harmonic oscillator potential. This
significant feature will be used later in this work to test our
results.

Using the spirit of the standard perturbation theory, we put
forward here a more general formalism to treat quantum mechanical
perturbation problems efficiently in a simple framework, which
has been already discussed in detail through the recent works \cite{gonulgspt, gonuletoellstates} involving
different applications of the model. Start with
\begin{equation}
\Psi _{n}(r)=\chi_{n}(r)\phi_{n}(r)~,
\end{equation}
in which $\chi_{n}$ is the known normalized eigenfunction for the exactly solvable
potential in (1) including the barrier term and $\phi_{n}$ is a
moderating function due to the perturbing potential. Substituting (3)
into (2) yields
\begin{equation}
\frac{\hbar ^{2}}{2m}\left(\frac{\chi''_{n}}{\chi_{n}}
+\frac{\phi''_{n}}{\phi_{n}}+2\frac{\chi'_{n}}{\chi_{n}}
\frac{\phi'_{n}}{\phi_{n}}\right)=V-E_{n}~.
\end{equation}

With the new definitions,
\begin{equation}
W_{n}=-\frac{\hbar}{\sqrt{2m}}\frac{\chi'_{n}}{\chi_{n}}~,~
\Delta W_{n}=-\frac{\hbar}{\sqrt{2m}}\frac{\phi'_{n}}{\phi_{n}}
\end{equation}
one arrives at
\begin{equation}
\frac{\hbar^{2}}{2m}\frac{\chi''_{n}}{\chi_{n}}=W_{n}^{2}
-\frac{\hbar}{\sqrt{2m}}W'_{n}=V_{ES}-\epsilon_{n}~,
\end{equation}
where $\epsilon _{n} $ is the eigenvalue for the exactly
solvable potential of interest, and
\begin{equation}
\frac{\hbar^{2}}{2m}\left(\frac{\phi''_{n} }{\phi _{n} }
+2\frac{\chi ' _{n} }{\chi _{n} } \frac{\phi ' _{n} }{\phi _{n} } \right)
=\Delta W_{n} ^{2} -\frac{\hbar }{\sqrt{2m} } \Delta W' _{n} +2W_{n}
\Delta W_{n} =\Delta V-\Delta \epsilon _{n}~,
\end{equation}
in which $\Delta \epsilon _{n}$ is the correction term to the energy due to
$\Delta V$, and $E_{n}=\epsilon_{n}+\Delta \epsilon_{n}$. Subsequently,
Eq. (4) reduces to
\begin{equation}
\left( W_{n} +\Delta W_{n} \right) ^{2} -\frac{\hbar }{\sqrt{2m} } \left(
W_{n} +\Delta W_{n} \right)'=V-E_{n}~.
\end{equation}

As the complete spectrum and wave functions are known in the
analytical form for the exactly solvable potentials appeared
in (6), one needs here to solve (7) to obtain in a closed form
the corrections to both, energy and wave function.

Proceeding with the perturbed Coulomb potential in arbitrary
dimensions,
\begin{equation}
V(r)=\left[ -\frac{a}{r} +\frac{\Lambda \left( \Lambda +1\right)
}{2mr^{2} } \right] +br+cr^{2},
\end{equation}
and set the superpotential, for the ground solutions,
\begin{equation}
W_{n=0} (r)=\sqrt{\frac{m}{2} } \frac{a}{\left( \Lambda +1\right) \hbar }
-\frac{\left( \Lambda +1\right) \hbar }{\sqrt{2m} r}~,
\end{equation}
leading to the first part of the potential in the bracket, and
from the literature, corresponding normalized wave function and
energy are
\begin{equation}
\chi _{n=0} (r)=N_{0} r^{\Lambda +1} \exp \left[ -\frac{ma}{\left(
\Lambda +1\right) \hbar ^{2} } r\right]~,~\epsilon _{n=0}
=-\frac{ma^{2} }{2\hbar ^{2} \left( \Lambda +1\right) ^{2} }~.
\end{equation}

For the perturbing potential, $\Delta V=br+cr^{2} $, the
physically acceptable unique choice is
\begin{equation}
\Delta W_{n=0} (r)=\sqrt{c} r~,
\end{equation}
which satisfies Eq. (7) from where one readily sees that
\begin{equation}
\Delta \epsilon _{n=0} =\frac{M\left( M-1\right) b\hbar ^{2} }{4ma}~,~
b=\frac{2a\sqrt{2mc} }{\left( M-1\right) \hbar }~.
\end{equation}

It is stressed that such solutions in general have constraints
on the potential parameters as appeared in (13). These constraints
differ for each eigenvalue, and hence various solutions do not
correspond to the same potential and are not orthogonal.

From (12), the moderating function
\begin{equation}
\phi _{n=0} (r)=\exp \left[ -\frac{\sqrt{2m} }{\hbar }
\int\limits_{}^{r}\Delta W_{n=0} (r) \right] =\exp \left[ -\frac{b\left(
M-1\right) }{4a} r^{2} \right]~.
\end{equation}
Therefore, the full wave function $\Psi_{n=0}$ for the potential in (9) takes the form
\begin{equation}
\Psi _{n=0} (r)=\chi _{n} (r)\phi _{n=0} (r)=N_{0}~r^{\left(
\frac{M-1}{2} \right) } \exp \left[ -\frac{2ma}{\left( M-1\right) \hbar
^{2} } r-\frac{b\left( M-1\right) }{4a} r^{2} \right]~.
\end{equation}

Finally, the exact ground state energy is
\begin{equation}
E_{n=0} =\epsilon _{n=0} +\Delta \epsilon _{n=0} =
-\frac{2ma^{2} }{\hbar ^{2} \left( M-1\right) ^{2} } +\frac{M\left(
M-1\right) b\hbar ^{2} }{4ma}~.
\end{equation}
The results in Eqs. (15) and (16) are exact and agree with
those in \cite{chaudhuri18501995, morales8631999}.

As noted earlier, the potential in (9) behaves also like a harmonic
oscillator in case for large $c$ and relatively small $a$ and $b$. This
enables one to check out explicitly the results obtained
above and the reliability of the formalism introcuded. For this
consideration, we set the superpotentials
\begin{equation}
W_{n=0} (r)=\sqrt{c} r-\frac{\left( \Lambda +1\right) \hbar }{\sqrt{2m}
r}~,~\Delta W_{n=0} (r)=\frac{b/2}{\sqrt{c}}~,
\end{equation}
which yield, through the use of either Eqs. (6) and (7) together,
or (8) alone,
\begin{equation}
\epsilon _{n=0} =\frac{\hbar \sqrt{c} }{\sqrt{2m} } \left( 2\Lambda
+3\right)~,~\Delta \epsilon _{n=0} =-\frac{b^{2} }{4c}~,
\end{equation}
which are exactly equivalent to those found for the perturbed
Coulomb case in (16). Similarly, using (17) one readily arrives
at (15).

For the generalization, we further make clear that the present
technique is also applicable to excited states, for which one
needs to use the shape invariance property and the relation between
supersymmetric partners \cite{cooper2671995}
\begin{equation}
V^{+} (r,\alpha _{0} )=V^{-} (r,\alpha _{1} )+R(\alpha _{1} )~,
\end{equation}
where $\alpha _{0} =\Lambda $ and
$\alpha _{1} =f\left( \alpha _{0} \right) =\Lambda +1$ are the
position independent parameters while $V^{\pm } $ are the supersymmetric partners
\begin{equation}
\left( W_{n=0} +\Delta W_{n=0} \right) ^{2} \mp \frac{\hbar }{\sqrt{2m} }
\left( W_{n=0} +\Delta W_{n=0} \right)' =V^{\mp } -E_{n=0}^{-}~,
\end{equation}
which depend upon superpotentials in (10) and (12) for the case $a>b,c$.
Bearing in mind that the replacement of $\Lambda $ with $\Lambda +1$ in
$V^{-} (r,\Lambda )$ leads to $V^{-} (r,\Lambda +1)$ and keeping
the potential parameters $b$ and $c$ fixed as in \cite{chaudhuri18501995, morales8631999} while allowing the change in
$a$, see Eq. 13, we obtain by the straightforward calculations
\begin{eqnarray}
E_{n}&=&E_{n}^{-} -\frac{b^{2} }{4c} +\frac{\hbar \sqrt{c} }{\sqrt{2m} }
\left( 2\Lambda +3\right)~,~
\nonumber
\\
E_{n}^{-}&=&\sum\limits_{k=1}^{n}R(\alpha _{k} )=-\frac{\hbar \sqrt{c} }{\sqrt{2m} }
\left[ \left( 2\Lambda +3\right) -2\left( n+\Lambda \right) +3\right]~,~
\nonumber
\\
E_{n}&=&-\frac{b^{2} }{4c} +\frac{\hbar \sqrt{c} }{\sqrt{2m} } \left[
2\left( n+\Lambda \right) +3\right]~,~n=0,1,2,\ldots ~,
\end{eqnarray}
which agrees with \cite{morales8631999} in $N$-dimensional space and
with \cite{roychoudhury30251988} in three dimensions. Furthermore, one
can easily construct the bound $n^{th} $
state wave functions from the ground state wave function using
the supersymmetric definition \cite{cooper2671995}
\begin{equation}
\Psi _{n+1}^{-} (r,\alpha _{0} )\propto A^{+} (r,\alpha _{0} )\Psi
_{n}^{-} (r,\alpha _{1} )~,~
A^{+} (r,\alpha _{0} )=-\frac{\hbar}{\sqrt{2m} }\left[ \frac{d^{2} }{dr^{2} } +
\frac{\Psi'_{n=0}(r,\alpha _{0} )}{\Psi _{n=0} (r,\alpha _{0} )} \right]~.
\end{equation}

We conclude with two remarks. First, the present formalism can
be generalized to all the polynomial forces
\begin{equation}
V_{m} (r)=Ar^{2m} +Br^{2m-1} +\ldots +F/r +G/r^{2}
\end{equation}
as an alternative but simple treatment to the other works \cite{znojil1571999},
and the references therein. With the consideration of the perturbed
Coulomb problem we have clarifed that the explicit solution of the
related Schr\"{o}dinger equation with the potential family in (23)
remains feasible in an almost complete parallel with their $m=1$
predecessors. Along this line, the works are in progress.

The second remark we wish to make is that apart from the inherent
interest one has in the existence of exact solutions, the results
reported here are likely to be useful in perturbation calculations
for the excited state energies and wave functions, particularly
if the method of other perturbation theories appeared in the
literature for evaluating second- and higher-order corrections
can be extended to cover such cases. In this context, we believe
that the simple form of our new approach to such problems opens
a new direction of development towards many practical applications
yet to be constructed and appreciated.

\end{document}